\documentclass[aps,prb,twocolumn,final,showpacs,amsmath,amssymb]{revtex4}
\newcommand{\figwidth}{0.99\linewidth}

\usepackage{graphicx}
\usepackage{bm}

\begin{document}

\title{Spin-dependent dipole excitation in alkali-metal nanoparticles}

\author{Yue~Yin} 
\author{Paul-Antoine~Hervieux} 
\author{Rodolfo~A.~Jalabert}
\author{Giovanni~Manfredi} 
\author{Emmanuel~Maurat}
\author{Dietmar~Weinmann}

\affiliation{Institut de Physique et Chimie des Mat{\'e}riaux de Strasbourg,
UMR 7504, CNRS-UdS,\\
23 rue du Loess, BP 43, 67034 Strasbourg Cedex 2, France}
\date{\today}

\begin{abstract}
We study the spin-dependent electronic excitations in alkali-metal
nanoparticles. Using numerical and analytical approaches, we focus on the 
resonances in the response to spin-dependent dipole fields. In the 
spin-dipole absorption spectrum for closed-shell systems, we investigate 
in detail the lowest-energy excitation, the ``surface paramagnon" predicted 
by L.\ Serra \textit{et al.} [Phys. Rev. A \textbf{47}, R1601 (1993)]. 
We estimate its frequency from simple assumptions for the dynamical 
magnetization density. In addition, we numerically determine the dynamical 
magnetization density for all low-energy spin-dipole modes in the spectrum.
Those many-body excitations can be traced back to particle-hole excitations 
of the noninteracting system. Thus, we argue that the spin-dipole modes are 
not of collective nature. In open-shell systems, the spin-dipole response 
to an electrical dipole field is found to increase proportionally with the 
ground-state spin polarization.
\end{abstract}


\maketitle

\section{Introduction} 
\label{sec:intro}

The optical absorption of small metal particles is dominated by the surface 
plasmon resonance \cite{kreibig-book,deHeer93,brack93,brechi92-93}. In this
collective excitation, the center-of-mass of the electrons moves back and 
forth with respect to the positive background leading to an oscillating charge
dipole. Pump-probe experiments using femtosecond laser pulses have
been widely used to study the relaxation of these excitations
\cite{bigot95,sun94}. Within the surface plasmon excitation all the
electrons oscillate in phase, irrespective of their spin. Thus,
its study only yields information on the charge dynamics. In
order to address the spin dynamics of nano-objects, time-resolved
magneto-optical Kerr effect measurements have been performed recently,
yielding the full trajectory of the magnetization in real space
for optically excited superparamagnetic nanoparticles \cite{andrade07}.

A substantial interest in the theoretical description of spin dynamics in 
nano-objects was aroused by the work of Serra \textit{et al.} \cite{serra93},
who found that a strong peak in the spin dipole absorption spectrum of 
alkaline nanoparticles exhausts a large fraction of the energy-weighted sum 
rule and drew the conclusion that this peak corresponds to a collective spin 
mode. In this excitation named ``surface paramagnon'', the spin degree of 
freedom appears in a crucial manner. While the surface plasmon can be excited 
by a dipole electric field, the surface paramagnon results when an excitation 
acts differently for spin up and spin down electrons. Such a field can be 
realized experimentally through the magnetic field component of 
electromagnetic waves having a wavelength considerably longer than the size 
of the system \cite{serra97}.

For the case of spherically symmetric systems with zero total spin, in the 
spin-dipole mode the center-of-mass of the electron system does not move. 
Therefore, such an excitation does not couple to an electric field in
closed-shell systems. Since transitions induced by electric dipole fields are 
the dominant mechanism, the surface paramagnon is difficult to observe in the 
optical absorption spectrum of such nanoparticles. However, for nanoparticles 
with no spherical symmetry and/or open-shell electronic systems, the spin 
dipole couples to electric dipole fields \cite{mornas96,kohl99,torres00}.
For relatively small systems, it has been concluded from calculations of the 
time-evolution of strong excitations that the coupling between the charge and 
the spin-modes is not crucially modified when the excitation strength is 
increased into the non-linear r{\'e}gime \cite{mornas96}. In systems that 
lack spherical symmetry scissor modes exist which can also be coupled to the 
spin-modes \cite{reinhard02}.

The difficulty to detect the spin-dipole mode in rotationally invariant
nano-objects becomes less restrictive when one studies semiconductor quantum
dots instead of metallic nanoparticles. For the typical sizes and electronic
densities of quantum dots the energies of the charge and spin dipole
excitations are of the order of a few meV. Therefore Raman scattering of 
visible light can be used (with selection rules depending on the polarization 
geometry \cite{pinczuk89}) to detect and study both kinds of excitations 
\cite{schueller98}. Finally, the strong electronic confinement of disc-shaped 
quantum dots in semiconductor hetero-structures results in very sharp 
resonances for dipolar excitations. The enhanced damping of the lowest
spin-dipole mode in the presence of a weak magnetic field that splits the 
single-particle excitations has been taken as an indication of the collective 
character of the lowest spin-dipole mode \cite{schueller98}.

The availability of experimental data has motivated theoretical work on the 
charge and spin density excitations in semiconductor few-electron quantum 
dots (see for example Ref.\ \onlinecite{lipparini-book}). In particular, the 
induced magnetization density has been studied \cite{serra97} and a 
correspondence between the spin-dipole modes and single-particle excitations 
has been observed numerically \cite{serra99}. The small deviation of the spin
density resonance energies from single-particle excitation energies has been 
explained by the absence of long-range Coulomb interaction terms in the 
energy of the spin modes. In addition, the excitation energies have been 
studied using the Quantum Monte Carlo technique \cite{colletti07} and within 
a semicalssical approach for the time-dependent charge and spin density
oscillations \cite{puente00}.

In this work, we use analytical (mean-field) and numerical time-dependent
local spin-density approximation (TDLSDA) approaches to study in detail the 
physics of the spin-dipole modes in alkali-metal particles. We address the 
important questions related to the specificity of the lowest frequency 
resonance as compared with the other excitations by identifying the modes 
with the corresponding dynamical magnetization densities.

By studying the evolution of the spin modes with the interaction we provide 
clear-cut arguments in the discussion over the collective \textit{versus} 
single-particle nature of the surface paramagnon. In addition we obtain the 
size scaling of the lowest resonance frequency and relate it with spill-out 
effects. In the case of open-shell systems we analyze the nature of the 
different spin-dipole excitations and the relationship with the charge modes.

While our numerical results are worked out for the case of alkaline metal
nanoparticles, most of our general conclusions hold for a broad class of
nanosystems, including semiconductor quantum dots.

The paper is organized as follows. In Sec.\ \ref{sec:model-approach}, we
describe the model for the electron dynamics in metal nanoparticles and the
numerical method we employ. We also present the computed spin-dipole 
absorption spectrum and the corresponding dynamical magnetization density for 
a typical example of a closed-shell system. In Sec.\ \ref{sec:spin-dip}, we 
use a phenomenological approach to describe the energetically lowest spin 
dipole excitation and derive results for its frequency based on plausible 
assumptions on the dynamical magnetization density. We compare the 
numerically obtained frequencies with the phenomenological ones resulting 
from increasingly accurate descriptions of the electronic dynamics. In 
Sec.\ \ref{sec:ph}, we present numerical results for the full absorption 
cross section in the frequency regime below the surface plasmon frequency. 
We follow the evolution of the absorption spectrum with the strength of
electron-electron interactions and find a one-to-one correspondence of the 
spin-dipole modes with the particle-hole excitations. In Sec.\ \ref{sec:open}, 
we study open shell clusters and discuss the possibility to observe the 
spin-dipole modes in the electric dipole absorption spectrum. We provide our 
conclusions in Sec.\ \ref{sec:conclusions}. In the appendices we present the 
details of our LSDA parametrization and the calculations for the case of a 
non-uniform ground-state electron density.

\section{Numerical approach to spin-dipole excitations}
\label{sec:model-approach}

In our study of the electronic excitations of nanoparticles, we restrict
ourselves to the electronic degrees of freedom and describe the confining 
effect of the ionic background by a spherical jellium model with sharp 
boundaries. Such a simplification can be justified for not too small metal 
particles. Furthermore, we do not consider thermal effects and therefore 
choose to work at zero temperature.

We start by introducing the formalism underlying the numerical approach to 
the absorption cross section corresponding to spin-dependent excitations of 
our model nanoparticles. We follow the formulation of the TDLSDA 
\cite{rajagopal78} as it is presented in Ref.\ \onlinecite{lipparini-book}.
In this framework the electronic system is described in atomic units
($\hbar=m=e=4\pi\epsilon_0=1$) using the Kohn-Sham equations
\begin{equation}\label{eq:ks}
  i \frac{\partial}{\partial t} \phi^{\sigma}_k (\mathbf{r},t)=
    \left(-\frac{1}{2} \nabla^2 + V^{\sigma}_\mathrm{eff}(\mathbf{r},t)\right)
    \phi^{\sigma}_k(\mathbf{r},t)\, ,
\end{equation}
where $\phi^{\sigma}_k$ is the $k^\mathrm{th}$ Kohn-Sham wave-function
with the quantum number $\sigma=\{\uparrow, \downarrow\}$ describing spin
projection onto the $\hat{\mathbf{z}}$-axis. These wave-functions allow us
to define the spin-dependent electron densities  
\begin{equation}
  n^{\sigma}(\mathbf{r},t)=\sum_{k\; \mathrm{occ}}
     \left| \phi^{\sigma}_k(\mathbf{r},t) \right|^2 \, ,
\end{equation}
where the sum runs over the single-particle like Kohn Sham levels $k$ that 
contribute to the many-body density. The electron density, magnetization 
density, and spin polarization are obtained from $n^{\sigma}$, respectively, as
\begin{subequations}
\label{eq:definitions}
\begin{eqnarray}
n&=&n^{\uparrow}+n^{\downarrow} \, ,
\label{eq:eldensity}\\
m&=&n^\uparrow-n^\downarrow \, ,
\label{eq:spindensity}\\
\xi&=&m/n  \, .
\end{eqnarray}
\end{subequations}

The effective potential in the Kohn-Sham equations can be written as
\begin{equation}
  V^{\sigma}_\mathrm{eff} (\mathbf{r},t)=
  V_\mathrm{c}(\mathbf{r})+V_\mathrm{H}(\mathbf{r},t)
  +V^{\sigma}_\mathrm{xc}(\mathbf{r},t)
  +V^{\sigma}_\mathrm{ex} (\mathbf{r},t)\, ,
\end{equation}
where $V_\mathrm{c}$ represents the confinement due to the jellium background,
$V_\mathrm{H}$ is the Hartree potential, $V^{\sigma}_\mathrm{xc}$ is the
exchange-correlation potential and $V^{\sigma}_\mathrm{ex}$ stands for the
external perturbation. The local character of the approximation is reflected
by the choice
\begin{equation}\label{eqn:exch-corr}
  V^{\sigma}_\mathrm{xc} (\mathbf{r},t) =
  \left.\frac{\partial}{\partial n^{\sigma}}
  \left(n \epsilon_\mathrm{xc}(n^{\uparrow},n^{\downarrow})\right)
  \right|_{\genfrac{}{}{0pt}{1}{n^{\uparrow}=n^{\uparrow}(\mathbf{r},t)}
                               {n^{\downarrow}=n^{\downarrow}(\mathbf{r},t)}} 
  \, ,
\end{equation}
where $\epsilon_\mathrm{xc}(n^{\uparrow}, n^{\downarrow})$ stands for the 
exchange-correlation energy density for which we use the parametrization of 
Perdew and Zunger \cite{perdew81} reproduced in App.\ \ref{sec:app-perdew}.

Within linear response theory we write the density changes induced by the 
external perturbation as
\begin{equation}
  \delta n^{\sigma}(\mathbf{r},\omega) = \sum_{\sigma'}
         \int \mathrm{d}\mathbf{r}'
         \chi^{\sigma\sigma'}(\mathbf{r},\mathbf{r}',\omega)
         V^{\sigma'}_\mathrm{ex}(\mathbf{r}',\omega) \, ,
  \label{eqn:ind_n}
\end{equation}
where $V^{\sigma}_\mathrm{ex}(\mathbf{r},\omega)$ is the Fourier transform
of the time-dependent external potential, and the response functions
$\chi^{\sigma\sigma'}$ obey the Dyson equation
\begin{widetext}
\begin{equation}
  \chi^{\sigma \sigma'}(\mathbf{r},\mathbf{r}',\omega)=
  \chi_{0}^{\sigma\sigma'}(\mathbf{r},\mathbf{r'},\omega)
  + \alpha_\mathrm{c}\sum_{\sigma_1\sigma_2} \iint \mathrm{d}\mathbf{r_1} 
   \mathrm{d}\mathbf{r_2}
   \ \chi_{0}^{\sigma\sigma_1}(\mathbf{r},\mathbf{r_1},\omega)
    \left[ \frac{1}{|\mathbf{r_1} - \mathbf{r_2}|}
      + K_\mathrm{xc}^{\sigma_1\sigma_2}(\mathbf{r_1},\mathbf{r_2})\right]
      \chi^{\sigma_2\sigma'}(\mathbf{r_2},\mathbf{r}',\omega)\, ,
\label{eqn:dyson}
\end{equation}
where we introduced the parameter $\alpha_\mathrm{c}=1$ which will later 
allow us to modulate artificially the importance of the electron-electron 
interactions in model calculations. The kernel of Eq.~\ref{eqn:dyson} is given 
by
\begin{equation} \label{eq:kernel}
  K^{\sigma_1 \sigma_2}_\mathrm{xc} (\mathbf{r_1}, \mathbf{r_2}) =
  \left.\frac{\partial}{\partial n^{\sigma_2}}
  \left(V^{\sigma_1}_\mathrm{xc}(n^{\uparrow},n^{\downarrow})
  \right)\right|_{\genfrac{}{}{0pt}{1}{n^{\uparrow}=n^{\uparrow}(\mathbf{r}_1)}
                                {n^{\downarrow}=n^{\downarrow}(\mathbf{r}_2)}} 
  \delta\left(\mathbf{r_1}-\mathbf{r_2}\right)\, .
\end{equation}
The non-interacting response function is diagonal in the spin indices and
given by the density-density correlator
\begin{equation}
  \chi^{\sigma \sigma'}_0(\mathbf{r},\mathbf{r}',\omega ) =
  \delta_{\sigma,\sigma'} \sum_{jk \; \mathrm{occ}} 
   \phi^{\sigma *}_{j}(\mathbf{r})\phi^{\sigma}_{k}(\mathbf{r})
          \phi^{\sigma *}_{k}(\mathbf{r}')\phi^{\sigma}_{j}(\mathbf{r}')
    \left\{
    \frac{1}{\omega -(\varepsilon^{\sigma}_{k}-\varepsilon^{\sigma}_{j})+i\eta }
    -\frac{1}{\omega +(\varepsilon^{\sigma}_{k}-\varepsilon^{\sigma}_{j})+i\eta }
    \right\}
\end{equation}
which can be expressed in terms of the retarded Green functions 
\cite{lipparini-book,serra93}. We have chosen the imaginary part in the 
denominator that ensures the convergence as $\eta= 8\, \mathrm{meV}$ 
($=2.94\times 10^{-4}$ in atomic units). The robustness of the final result 
with respect to variations of this parameter has been checked.

Defining the spin-independent part of the external perturbation
$V_\mathrm{ex,n}= (V_\mathrm{ex}^\uparrow+V_\mathrm{ex}^\downarrow)/2$ and
its spin-dependent counterpart
$V_\mathrm{ex,m}= (V_\mathrm{ex}^\uparrow-V_\mathrm{ex}^\downarrow)/2$, the
response of the charge and magnetization densities $n$ and $m$ can be 
expressed in matrix form as
\begin{equation}
\left(\begin{array}{c}
      \delta n(\mathbf{r},\omega) \\
      \delta m(\mathbf{r},\omega)
      \end{array}\right)
= \int\mathrm{d}\mathbf{r}'
\left(\begin{array}{cc}
      \chi_\mathrm{nn}(\mathbf{r},\mathbf{r}',\omega)&
            \chi_\mathrm{nm}(\mathbf{r},\mathbf{r}',\omega)\\
      \chi_\mathrm{mn}(\mathbf{r},\mathbf{r}',\omega)&
            \chi_\mathrm{mm}(\mathbf{r},\mathbf{r}',\omega)\\
      \end{array}\right)
\left(\begin{array}{c}
      V_\mathrm{ex,n}(\mathbf{r}',\omega)\\
      V_\mathrm{ex,m}(\mathbf{r}',\omega)
      \end{array}\right)\, ,
\end{equation}
\end{widetext}
where the cross-correlations of the charge and spin channels are given by
\begin{subequations}
\label{eq:cross_correlation}
\begin{eqnarray}
\chi_\mathrm{nn/nm}&=&\chi^{\uparrow\uparrow}\pm\chi^{\uparrow\downarrow}
                     +\chi^{\downarrow\uparrow}\pm\chi^{\downarrow\downarrow}
                     \, ,     \\
\chi_\mathrm{mn/mm}&=&\chi^{\uparrow\uparrow}\pm\chi^{\uparrow\downarrow}
                     -\chi^{\downarrow\uparrow}\mp\chi^{\downarrow\downarrow}
                     \, .
\end{eqnarray}
\end{subequations}

Electromagnetic radiation with wavelength much larger than the size of the
nanoparticles induces dipolar perturbations. Considering monochromatic light 
with wave-vector $\mathbf{k}=k \hat{\mathbf{x}}$, linear polarization along 
$\hat{\mathbf{y}}$, and therefore magnetic field along the 
$\hat{\mathbf{z}}$-direction, the dipole excitation potentials for
charge and spin can be written as
\begin{subequations}
\label{eq:excitation}
\begin{eqnarray}
V_\mathrm{ex,n}(\mathbf{r},t)&=&F_\mathrm{n}\, y \, ,
\label{eq:excitation_c}
\\
V_\mathrm{ex,m}(\mathbf{r},t)&=&F_\mathrm{m}\, x \, ,
\label{eq:excitation_m}
\end{eqnarray}
\end{subequations}
with the excitation strengths $F_\mathrm{n}=-E_\mathrm{max}\sin(\omega t)$
and $F_\mathrm{m}=g\mu_\mathrm{B}E_\mathrm{max}k\cos(\omega t)$, the Bohr 
magneton $\mu_\mathrm{B}$, and the gyromagnetic factor $g$ ($g\mu_\mathrm{B}$=1 
in atomic units). These dipolar perturbations lead to dipolar charge and 
spin-density excitations, and the corresponding polarizabilities are given by
\begin{equation}\label{eq:alpha}
\alpha_\mathrm{ab}(\omega)=
       \iint \mathrm{d}\mathbf{r} \mathrm{d}\mathbf{r}'\, rr'
       \cos{\theta}\ \cos{\theta'} \
       \chi_\mathrm{ab}(\mathbf{r},\mathbf{r}',\omega)\, ,
\end{equation}
with ab=\{nn, nm, mn, mm\}. For the simplicity of notations, the polar 
coordinates ($r,\theta,\varphi$) are here and henceforth chosen to have the 
$\hat{\mathbf{z}}$-axis along the variation of the excitation field. This
conventional choice allows us to treat the charge and spin excitations
within the same description, but it is not consistent with the example
of electromagnetic radiation presented above. 

\begin{figure*}
  \includegraphics[width=8cm]{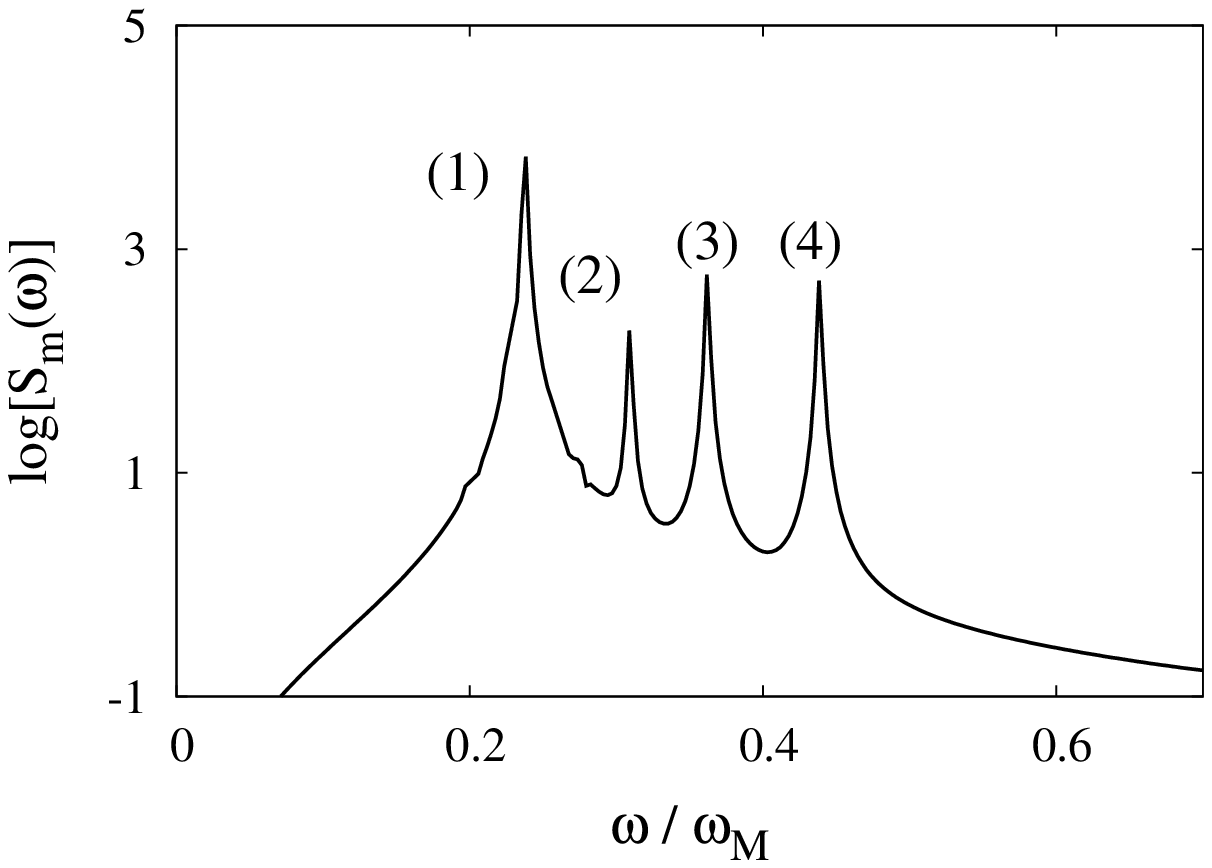}
  \hfill
  \includegraphics[width=8cm]{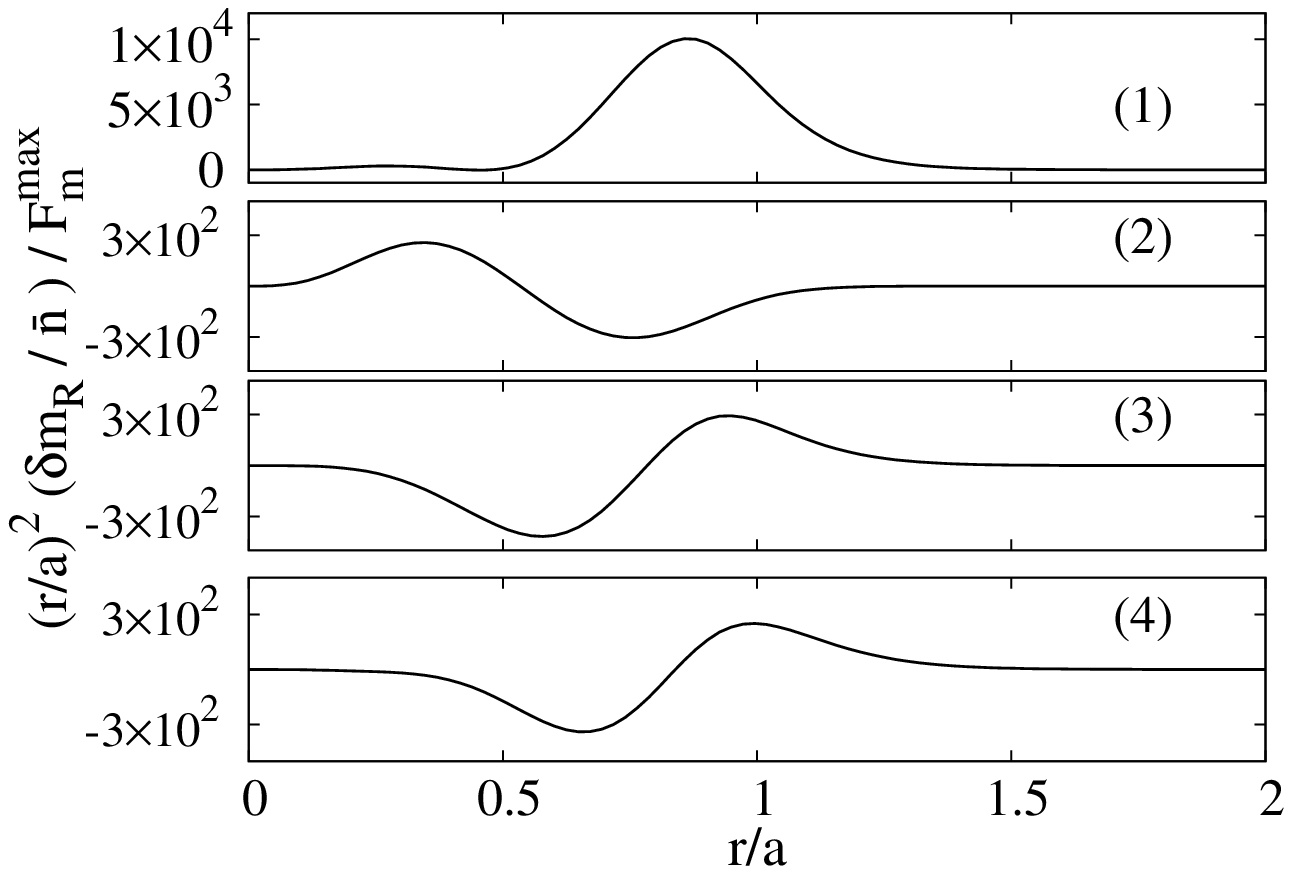}
  \caption{Left: spin-dipole absorption spectrum for a Na$_{34}$ 
    nanoparticle. The frequencies of the horizontal axis are normalized
    to the Mie frequency $\omega_{\rm M}=3.4\, \mathrm{eV}$ ($=0.125$ in 
    atomic units).
    Right: radial part of the magnetization density for each of the resonances 
    identified in the left panel, as a function of the radial coordinate
    (scaled with the radius $a$ of the particle). The magnetization profile
    is scaled with the excitation strength, the mean density and
    the radial coordinate}
  \label{fig:overviewNa34}
\end{figure*}

The experimentally relevant quantities are the dipole absorption cross-sections
\begin{equation}\label{eq:S}
S_\mathrm{ab}(\omega)= \frac{4\pi\omega}{c}
\mathop{\rm Im} \left[ \alpha_\mathrm{ab}(\omega) \right]\, .
\end{equation}
For spherically symmetric nanoparticles $\alpha$ and $S$ are diagonal in the
channel indices a and b. In this case we will work with
$S_\mathrm{n}=S_\mathrm{nn}$ and $S_\mathrm{m}=S_\mathrm{mm}$. The spin-dipole 
absorption spectrum $S_\mathrm{m}(\omega)$ for the closed-shell system
Na$_\mathrm{34}$ \cite{na34} is shown in the left panel of
Fig.\ \ref{fig:overviewNa34}. Four peaks are observed in the low-energy range 
below $0.6 \, \omega_\mathrm{M}$, with the Mie frequency 
$\omega_\mathrm{M}=3.4\,\mathrm{eV}$ which is the classical frequency of the 
surface plasmon excitation \cite{kreibig-book}. The peak at the lowest 
frequency, labeled (1) in the figure, displays the strongest absorption 
cross-section and corresponds to the surface paramagnon described by Serra
\textit{et al.} \cite{serra93}. In section \ref{sec:spin-dip} we derive 
analytical expressions that accurately describe its frequency and its 
dependence on the size of the nanoparticle.

The radial part of the magnetization density at resonance is shown in 
Fig.\ \ref{fig:overviewNa34} (right) for the four peaks appearing in the 
absorption spectrum. The magnetization profile for the lowest frequency peak 
(1) clearly differs from the profiles corresponding to the higher frequency 
peaks: it involves considerably stronger magnetization densities than the 
other peaks and, most importantly, it displays no significant nodes (except 
at the center of the nanoparticle), whereas the other peaks are associated 
to magnetization profiles with richer node structures inside the nanoparticle. 
For larger systems, even more peaks appear, and the corresponding 
magnetization profiles show more complicated structures with several nodes. 
However, the particularly simple structure of the lowest-frequency peak, and 
its stronger amplitude, persist and thereby point to its special character.

\section{Phenomenological approach to spin-dipole excitations}
\label{sec:spin-dip}

The physics of the spin-dipole excitations obtained in the previous section 
can be understood through phenomenological models. In particular, we present 
an estimation of the lowest resonant frequency and compare it with results 
from TDLSDA calculations.

We consider a spherically symmetric nanoparticle with ground-state equilibrium 
densities $n_0^{\uparrow}(\mathbf{r})=n_0^{\downarrow}(\mathbf{r})$ 
and a perturbation in the spin channel such that 
$\Delta n^{\uparrow}(\mathbf{r})=- \Delta n^{\downarrow}(\mathbf{r})$. 
The displacements of the center-of-mass of the two spin populations 
along the $\hat{\mathbf{z}}$-direction are given by
\begin{equation}\label{eqn:dev3}
Z^\uparrow = -  Z^\downarrow =  \frac{1}{N^\uparrow}
     \int \mathrm{d}\mathbf{r}\, z \ \Delta n^\uparrow(\mathbf{r})\, ,
\end{equation}
where $N^\uparrow=N/2$ is the number of spin-up electrons and $N$ the total
number of electrons. Since there is no net charge displacement, the Hartree 
term $E_\mathrm{H}$ of Eq.~\ref{eq:Hartree} remains unchanged under the
perturbation. The changes in the other contributions to the total energy 
can be calculated from $ \Delta n^\uparrow(\mathbf{r})$.

\subsection{Uniform ground-state density}
\label{subsec:ued}

The ground-state equilibrium electron density in a spherical jellium model 
of radius $a$ with sharp boundaries can be approximated by a uniform 
distribution
\begin{equation}\label{eqn:dev1}
 n^{\uparrow}(\mathbf{r})=n^{\downarrow}(\mathbf{r})
 =\frac{\bar{n}}{2} \
\Theta(a-r)
\end{equation}
inside the sphere, where $\bar{n}=3N/4\pi a^3$ and $\Theta$ denotes the 
Heaviside function. Assuming that the perturbation is a dipolar field, the 
simplest approximation to describe the low-energy spin excitations of the 
system is to postulate the tilts
\begin{equation}\label{eqn:dev2}
  \Delta n^{\uparrow}(\mathbf{r}) = - \Delta n^{\downarrow}(\mathbf{r}) =
  \frac{z}{\zeta} \ \frac{\bar{n}}{2} \ \Theta(a-r)
\end{equation}
of the spin densities. The characteristic length $\zeta$ describes the 
magnitude of the excitation. Working in linear response, we restrict 
ourselves to weak excitations with $\zeta \gg a$ and consider the change 
in total energy induced by the above density excitations using the energy 
functionals described in App.\ \ref{sec:app-perdew}. 

The simple form assumed for the spin densities allows us to neglect the change 
of $E_\mathrm{K,G}$. To the lowest order in the perturbation the other 
components of the energy are modified as
\begin{subequations}
\label{eq:energy_changes}
\begin{eqnarray}
  \Delta E_{\mathrm{K,TF}}&=&\frac{2\pi^{4/3}}{3^{1/3}} 
   \int \mathrm{d}\mathbf{r} \ \frac{(\Delta n^\uparrow)^2}{\bar{n}^{1/3}} 
  \nonumber \\ &=& \frac{5}{4}\genfrac{(}{)}{}{}{3\pi^2}{2}^{1/3} N^{5/3} \frac{Z^{\uparrow 2}}{a^4} \, , \\
  \Delta E_\mathrm{X}&=& - \frac{2}{(9\pi)^{1/3}} 
   \int \mathrm{d}\mathbf{r} \ \frac{(\Delta n^\uparrow)^2}{\bar{n}^{2/3}} 
  \nonumber \\ &=&  - \frac{5}{6}\genfrac{(}{)}{}{}{3}{2\pi}^{2/3} N^{4/3} \frac{Z^{\uparrow 2}}{a^3} \, , \\
  \Delta E_\mathrm{C}&=&  \frac{2}{(9\pi)^{1/3}} 
   \int \mathrm{d}\mathbf{r} \ c(\bar{r}_\mathrm{s}) \ 
   \frac{(\Delta n^\uparrow)^2}{\bar{n}^{2/3}} 
   \nonumber \\ &=& \frac{5}{6}\genfrac{(}{)}{}{}{3}{2\pi}^{2/3} c\left(\bar{r}_\mathrm{s}\right) N^{4/3}
   \frac{Z^{\uparrow 2}}{a^3} \, .
\end{eqnarray}
\end{subequations}
We have defined
\begin{equation}\label{eqn:cofr}
c\left(r_\mathrm{s}\right) =  \frac{1}{3}\left(\frac{2^{4/3}}{2^{1/3}-1}\right)
\left(\frac{4\pi}{3}\right)^{2/3} r_\mathrm{s} \left[\epsilon^{P}
\left(r_\mathrm{s}\right)-\epsilon^{U}\left(r_\mathrm{s}\right)\right]
\end{equation}
with $\epsilon^{P,U}$ given in App.\ \ref{sec:app-perdew} and 
$\bar{r}_\mathrm{s}=(4\pi\, \bar{n}/3)^{-1/3}$. The energy increase due to the
spin density displacements leads to a restoring force
$\mathcal{F}=-\partial E/\partial Z^\uparrow$ and an out-of-phase oscillation 
of the two spin subsystems with a frequency
\begin{eqnarray}\label{eqn:freq_mu2}
&&\omega_\mathrm{S} =\sqrt{\frac{2 \ \Delta E}{N \ Z^{\uparrow 2}}}
\\ \nonumber &&=
\frac{\omega_\mathrm{M}}{N^{1/3}}\genfrac{(}{)}{}{}{3}{2\pi}^{1/3}
\left[\frac{5}{3}
\left(\genfrac{(}{)}{}{}{3\pi^2}{2}^{2/3}\frac{1}{\bar{r}_\mathrm{s}} - 1 +
c(\bar{r}_\mathrm{s})\right)\right]^{1/2} .
\end{eqnarray}
We have expressed the result in terms of the classical Mie frequency 
\cite{kreibig-book} which can be written as 
$\omega_\mathrm{M} = \sqrt{N/a^3}=\left(\bar{r}_\mathrm{s}\right)^{-3/2}$. 

As compared to the spin dipole, the surface plasmon excitation is of quite 
different nature since it results from the oscillation of the total charge. 
The frequency $\omega_\mathrm{M}$ can be obtained following similar lines as 
those presented above, but restricting the restoring force to the Hartree 
contribution. The different nature of the energies involved in each mode 
results in a higher frequency for the surface plasmon (in the visible part of 
the spectrum for the case of metal nanoparticles) than for the spin dipole 
(in the infrared range). Moreover, $\omega_\mathrm{M}$ is independent of the 
size of the particle, while $\omega_\mathrm{S}$ decreases with the number of 
electrons as $N^{-1/3}$. This power-law scaling has already been obtained in 
Ref.\ \onlinecite{lipparini-califano96} within sum rule and hydrodynamic 
approaches. It makes the observation of the surface paramagnon in 
large particles more difficult.

\begin{figure}
\centerline{\includegraphics[width=\figwidth]{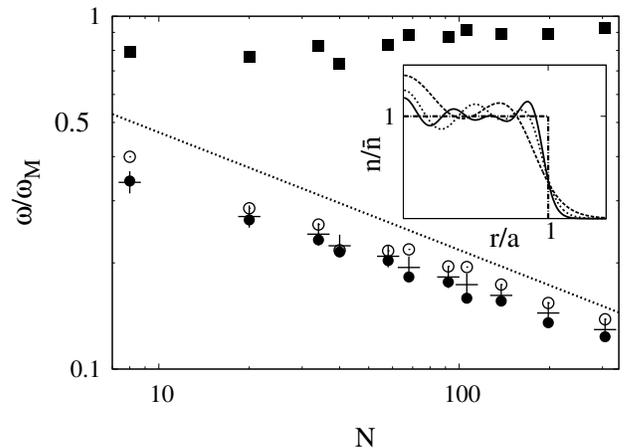}}
  \caption{Size-dependence of the surface paramagnon frequency, 
   together with the surface plasmon frequency 
   $\omega_\mathrm{sp}$ (filled squares).
   Filled circles represent $\omega_\mathrm{sd}$ obtained from TDLSDA
   calculations. The dotted line is the estimate $\omega_\mathrm{S}$ of 
   Eq.\ \ref{eqn:freq_mu2}. The open circles stand for the spill-out 
   corrected Eq.\ \ref{eqn:freq_mu2}, where $\omega_\mathrm{sp}$ is used 
   instead of $\omega_\mathrm{M}$ in the prefactor.
   The pluses depict the semi-analytical result of \eqref{eqn:freq_m}
   using the magnetization profile arising from Eq.\ \ref{eqn:in_mag5}.
   Inset: Radial variation of the ground-state electron density used in 
   Eq.~\ref{eqn:in_mag5}, and obtained from static LSDA calculations. The 
   dashed, dotted, and solid lines are for Na$_\mathrm{20}$, 
   Na$_\mathrm{106}$, and Na$_\mathrm{306}$, respectively.
   $\bar{n}$ is the electron density for bulk Na and the dash-dotted step 
   function corresponds to the uniform ionic jellium.}
  \label{fig:freq_N}
\end{figure}
In Fig.\ \ref{fig:freq_N} we compare the values of $\omega_\mathrm{sd}$
obtained from the TDLSDA (filled circles) with the estimate 
$\omega_\mathrm{S}$ of Eq.\ \ref{eqn:freq_mu2} (dotted line). For 
comparison we also show the numerically calculated surface plasmon 
frequencies $\omega_\mathrm{sp}$ (filled squares), which approach the 
classical value $\omega_\mathrm{M}$ for large $N$ and display important 
oscillations for small $N$ [\onlinecite{weick06}]. We can see that the predicted
decrease of the spin-dipole frequency as $N^{-1/3}$ is essentially correct. 
However, Eq.\ \ref{eqn:freq_mu2} overestimates the actual frequencies. This 
discrepancy becomes increasingly important when the size $a$ of the 
nanoparticle diminishes. Two key assumptions in the derivation of
\eqref{eqn:freq_mu2} become less justified when $a$ gets smaller. On one hand, 
the spill-out effect due to the extension of the electron wave-functions 
beyond the jellium sphere lowers the electron density as compared to the bulk 
value \cite{brack93,brechi92-93} (see Fig.\ \ref{fig:freq_N}, inset). On the 
other hand, assuming the tilt \eqref{eqn:dev2} and not considering density 
gradients in the energy functional may become problematic.

A simple way to approximately include spill-out effects is to use an
electronic density which is slightly lower than $\bar{n}$. In the case of
the surface plasmon, where the numerically obtained frequency 
$\omega_\mathrm{sp}$ is lower than $\omega_\mathrm{M}$, such an approach leads 
to a reduced frequency
$\tilde{\omega}_\mathrm{M}=\omega_\mathrm{M}\sqrt{1-N_\mathrm{out}/N}$, where
$N_\mathrm{out}$ is the number of electrons outside the jellium sphere.
However, $\omega_\mathrm{sp}$ is still lower than $\tilde{\omega}_\mathrm{M}$,
and moreover it exhibits a non-monotonous behavior not accounted for by
$\tilde{\omega}_\mathrm{M}$ (see Fig.~\ref{fig:freq_N} and
Ref.\ \onlinecite{weick06}). Assuming that the effect of spill-out on the
spin-dipole frequency is similar to the one on the surface plasmon frequency,
it is tempting to substitute $\omega_\mathrm{M}$ by $\omega_\mathrm{sp}$ in
Eq.\ \ref{eqn:freq_mu2}. As shown in Fig.\ \ref{fig:freq_N}, such an approach
(circles) considerably improves the estimation of $\omega_\mathrm{sd}$.

\subsection{Non-uniform ground-state density}
\label{subsec:nued}

A further improvement of the accuracy can be achieved by going beyond the 
approximation of the tilt \eqref{eqn:dev2} of the spin up and down densities 
and, at the same time, taking into account the spatial variations of the 
ground-state electron density. The latter consideration is crucial since in 
the spill-out region the density falls rapidly to zero (see inset of 
Fig.~\ref{fig:freq_N}), such that the kinetic energy contribution 
$E_\mathrm{K,G}$ of \eqref{eqn:ekingrad}, which includes the gradients of the 
electronic densities, becomes important.

In this section, we assume that the magnetization profile of the surface 
paramagnon is given by the static magnetization induced by a static external 
dipolar magnetic field. Expressing the energy functional of 
Eq.\ \ref{eqn:eng1} in terms of the charge and magnetization densities, the 
ground-state conditions for $n_0(\mathbf{r})$ and $m_0(\mathbf{r})$ are
\begin{equation}\label{eqn:eq_cond}
  \left.\frac{\delta E[n,m]}{\delta n(\mathbf{r})}
  \right|_{\genfrac{}{}{0pt}{1}{n=n_0(\mathbf{r})}{m=m_0(\mathbf{r})}} = 
  \left.\frac{\delta E[n,m]}{\delta m(\mathbf{r})}
  \right|_{\genfrac{}{}{0pt}{1}{n=n_0(\mathbf{r})}{m=m_0(\mathbf{r})}}
  =0 \, .
\end{equation} 

Applying an external magnetic field along the $z$ axis, 
$\mathbf{B}_\mathrm{ex}(\mathbf{r})={B}_\mathrm{ex}(\mathbf{r})\hat{\mathbf{z}}$ 
results (with $g\mu_\mathrm{B}=1$ in atomic units and the negative charge 
of the electron) in an additional contribution to the total energy functional
\begin{equation}
E_{T}[n,m]=E[n,m]+ 
\int \mathrm{d}\mathbf{r} \ B_\mathrm{ex}(\mathbf{r}) \ m(\mathbf{r})\, .
\end{equation}

If we work with a spherically symmetric nanoparticle, the charge and spin 
channels are decoupled. Thus, in linear response, the application of a 
magnetic field does not affect $n_0(\mathbf{r})$, and we drop this 
functional variable hereafter. The magnetization density is driven from its 
ground-state value $m_0(\mathbf{r})=0$ to a perturbed value 
$\Delta m(\mathbf{r})$, which is given by 
\begin{equation}
\left. \frac{\delta E[m]}
{\delta m(\mathbf{r})}\right|_{m=\Delta m(\mathbf{r})} = 
-B_\mathrm{ex}(\mathbf{r}) \, . 
\label{eq:magnet}
\end{equation}

Once the applied field is removed, the nanoparticle is left with an extra-energy
\begin{eqnarray}
\Delta E &\simeq& \frac{1}{2} \int \mathrm{d}\mathbf{r}\int \mathrm{d}\mathbf{r'} 
\left.\frac{\delta^2 E[m]}{\delta m(\mathbf{r}) \delta m(\mathbf{r'})}
\right|_{m=m_0} \Delta m(\mathbf{r})\Delta m(\mathbf{r'}) 
\nonumber \\
&=&  \frac{1}{2} \int \mathrm{d}\mathbf{r} \ B_\mathrm{in}(\mathbf{r}) 
   \Delta m(\mathbf{r}) \, .
\label{eq:ener_quad}
\end{eqnarray}
In the last equality we have used the perturbed equilibrium condition
(\ref{eq:magnet}) and defined $B_\mathrm{in}=- B_\mathrm{ex}$ as an internal 
field that counterbalances the applied one. Once the perturbation is switched 
off, the dynamics of the magnetization is determined by the excess energy 
$\Delta E$.

For the dipolar excitations that we are interested in (\textit{i.e.} 
Eq.~\ref{eq:excitation_m}) an appropriate choice for the external field is 
$B_\mathrm{ex}=- z/\lambda_B$, with $1/\lambda_B$ measuring the strength of 
the perturbation. With \eqref{eqn:dev3}, this allows us to write the extra 
energy as
\begin{equation}
  \Delta E = \frac{1}{2\lambda_B}\int \mathrm{d}\mathbf{r} \ z \ 
  \Delta m(\mathbf{r})
  = \frac{N Z^{\uparrow}}{2 \lambda_B} \, .
  \label{eqn:en_incr}
\end{equation}
The condition \eqref{eq:magnet} and the form of $B_\mathrm{ex}$ fix the 
induced magnetization, which we can write as 
$\Delta m(\mathbf{r})= \delta m_\mathrm{R}(r) \cos\theta$. Defining the
quantity $\tilde{m}(r)=\lambda_B r^2 \delta m_\mathrm{R}(r)$, we show in 
App.~\ref{sec:app-calc_nu} that in the linear r\'egime the magnetization 
profile is determined by the differential equation 
\begin{widetext}
\begin{equation}
  -D(r) \ \tilde{m}^{\prime}(r) +
  \left[ D(r)\left( \frac{2}{r} +36 D(r) \right) + A_\mathrm{KS}(r)\right]
  \tilde{m}(r) = n_0(r)r^3\, ,
  \label{eqn:in_mag5}
\end{equation}
where
\begin{subequations}
\label{eq:def_dif_eq}
\begin{eqnarray}
  \label{eq:def_dif_eq_a}
  A_\mathrm{KS}(r) & = & 
         \genfrac{(}{)}{}{}{1}{12\pi^2}^{1/3}\frac{1}{r_\mathrm{s}(r)}
         \left[\genfrac{(}{)}{}{}{3\pi^2}{2}^{2/3}\frac{1}{r_\mathrm{s}(r)}
         -1+c(r_\mathrm{s}(r))\right] 
  \, , \\
  \label{eq:def_dif_eq_b}
  D(r) & = & \frac{1}{36} \ \frac{n_0^{\prime}(r)}{n_0(r)} \, ,
\end{eqnarray}
\end{subequations}
\end{widetext}
with $n_0(r)$ being the electron density in the ground state which can be 
calculated numerically from a static LSDA code. The primes denote derivatives 
with respect to $r$.

\begin{figure}
  \includegraphics[width=\figwidth]{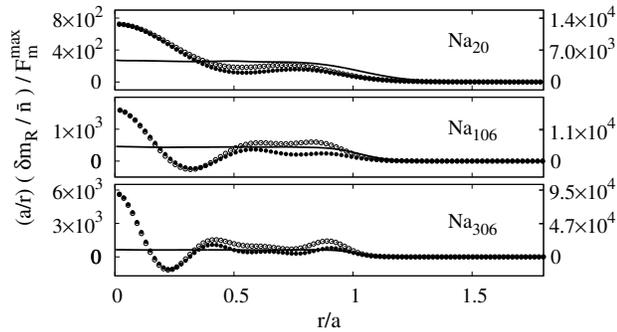}
  \caption{Radial part of the magnetization profile for various particle 
    sizes. The filled circles represent the TDLSDA calculations near the 
    frequency of the surface paramagnon (right scale). The empty circles 
    are the static results obtained from LSDA with an external field 
    $B_\mathrm{ex}=- z/\lambda_B$ (left scale). The solid line is the 
    solution of \eqref{eqn:in_mag5} based on a non-uniform ground-state
    electron density. The normalization of the magnetization profile
    differs from that of Fig.~\ref{fig:overviewNa34} by a factor 
    $(r/a)^3$, such that the tilted spin densities \eqref{eqn:dev2}
    result in a step function.}
  \label{fig:in_mag}
\end{figure}
As in the simpler case of a tilt-like magnetization, we assume that the 
functional form of the magnetization profile is conserved, up to an overall 
factor, in the oscillations occurring when the external field is switched off.
Such an assumption is supported by the numerical results shown in 
Fig.\ \ref{fig:in_mag}. The magnetization profile obtained at resonance 
($\omega=\omega_\mathrm{sd}$, filled circles, right scale) is much stronger but 
very close in shape to the static one 
($\omega=0$, $B_\mathrm{ex}=- z/\lambda_B$, empty circles, left scale). 
It is important to notice that the solution $\tilde{m}(r)$ of 
Eq.\ \ref{eqn:in_mag5} (thin solid line) is a good representation of the 
local spin-density calculations. The differences for small values of $r$ are 
not significant because of the volume integrals that are performed. In 
addition, we see from Fig.\ \ref{fig:in_mag} that the various approximations 
for the magnetization profile do not deviate considerably from the simple tilt 
\eqref{eqn:dev2} that we used in the previous chapter. Even if the 
magnetization profile attains its maximum value around $r=a$, the spin dipole 
is not a surface mode (in contrast to the surface plasmon), since the 
excitation is not confined to the surface but appears in the whole 
nanoparticle. 

The restoring force associated with $\Delta E$, Eq.\ \ref{eqn:en_incr}, 
leads to oscillations of the two spin populations with a frequency
\begin{equation}
\hat\omega_\mathrm{S} = \sqrt{\frac{2 \ \Delta E}{N \ Z^{\uparrow 2}}} =
  \sqrt{\frac{3}{4\pi} \frac{N}{\int \mathrm{d}r \, r \tilde{m}(r)}}\, .
  \label{eqn:freq_m}
\end{equation}
Using the profile $\tilde{m}(r)$ from (\ref{eqn:in_mag5}) we obtain a good 
approximation (pluses in Fig.~\ref{fig:freq_N}) of the numerically obtained 
$\omega_\mathrm{sd}$. This shows the importance of the spill-out in 
determining the frequency of the spin-dipole excitations, and underlines that 
the corresponding shift can be accurately estimated from the equilibrium 
density profiles \cite{note-sp}.

\section{Spin-dipole spectrum and particle-hole excitations} 
\label{sec:ph}

In the preceding section, we have identified the behavior of the lowest 
frequency peak in the spin-dipole absorption spectrum. Two important questions 
deserve to be addressed now. The first concerns the specificity of the lowest 
energy peak as compared with the other ones. The second question, already 
treated in the literature \cite{serra93}, partly in the context of electronic 
excitations in quantum dots \cite{serra99}, is whether or not the spin dipole 
can be considered to be a collective excitation, as it is the case for the
surface plasmon.

In Fig.\ \ref{fig:overviewNa34} we saw that in addition to its considerably 
larger oscillator strength, the first peak is peculiar from the point of view 
of the induced magnetization which has significant contributions of constant 
sign, whereas there are always important contributions of different sign for 
the other peaks. The assumption of a tilt for the magnetization profile used 
in our analytical approach of Sec.\ \ref{subsec:ued} is appropriate to 
describe a mode without nodes and thus allows for an estimate of the 
frequency of the lowest peak. In the semi-analytical model of 
Sec.\ \ref{subsec:nued}, the magnetization is supposed to be generated by an 
external magnetic field that is linear in $z$. The radial components of the 
magnetization profiles obtained numerically are also positive except for a 
very small insignificant region close to the center in the largest particles 
(see Fig. \ref{fig:in_mag}). In order to predict the frequencies of 
higher-energy peaks in the spectrum, one would have to assume more complex 
magnetization profiles. However, those peaks show no obvious regularities in 
their magnetization profile (\textit{e.g.} the number of nodes does not 
increase monotonically when moving to higher frequencies). One can conclude 
that the first peak really stands out as the only one with an essentially 
everywhere-positive magnetization, a property that allowed us to construct a 
rather precise theory for the frequency associated with this mode.

The authors of Ref.\ \onlinecite{serra93} concluded that the surface
paramagnon is a collective excitation, based on the observation that the 
resonance exhausts more than 90\,\% of the total spectral weight. 
\begin{figure}
  \includegraphics[width=\figwidth]{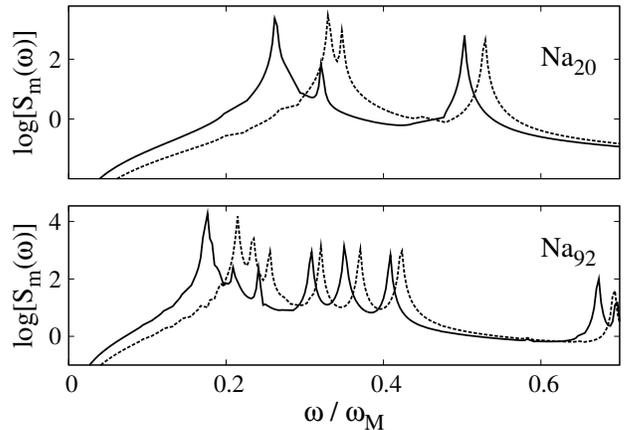}
  \caption{Spin-dipole absorption spectrum (solid line) and particle-hole 
    excitation spectrum (dashed line) for two nanoparticle sizes.}
  \label{fig:phtdlsda}
\end{figure}
A widely accepted criterion is to consider an excitation as collective if it 
results from the superposition of a large number of low-energy particle-hole 
excitations. This is certainly the case of the well-studied surface plasmon 
\cite{yannouleas92,seoanez07}, where the residual interaction (understood in 
this context as going beyond the Hartee-Fock approximation) results in a 
small perturbation of most of the particle-hole excitations and the 
appearance of the collective excitation in the high-energy sector of the 
spectrum. The considerably lower energy of the spin dipole and its size 
scaling suggest some important differences in the nature of the two 
excitations. In order to test this conjecture, we show in 
Fig.\ \ref{fig:phtdlsda} a comparison between the full spin-dipole absorption 
spectrum (solid lines) and the corresponding particle-hole excitation 
spectrum (dashed). The latter is obtained by removing the electron-electron 
interaction in the calculation of the linear response (setting 
$\alpha_\mathrm{c}=0$ in Eq.\ \ref{eqn:dyson}), although it is still included 
when computing the ground state. Since there is no Hartree contribution for 
the spin modes, the residual interaction in this context is understood as the 
effective exchange-correlation term \eqref{eq:kernel}. We can see that the two 
spectra have similar structure, with a one-to-one correspondence between the 
excitations. The full spin-dipole absorption spectrum appears to be slightly 
red-shifted as compared to the particle-hole spectrum because of the 
attractive nature of the exchange-correlation interaction.

\begin{figure}
  \includegraphics[width=\figwidth]{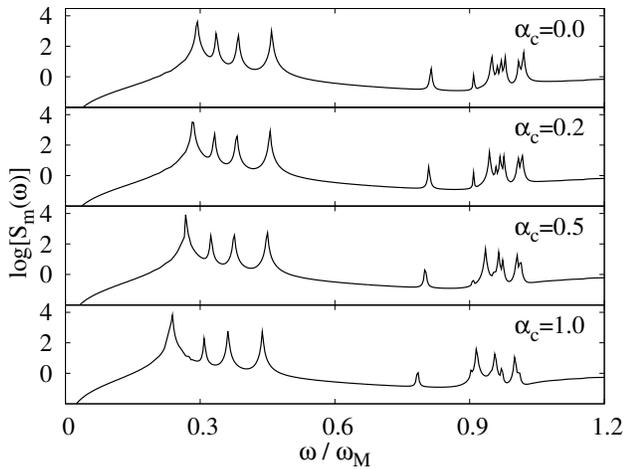}
  \caption{Spin-dipole absorption spectrum for Na$_\mathrm{34}$ and increasing 
   values of $\alpha_\mathrm{c}$ corresponding to an increasing importance of 
   the interactions.}
  \label{fig:eph_m}
\end{figure}
The selection rules for dipole-created electron-hole excitations dictate a 
minimal absorption energy \cite{seoanez07}, associated with a frequency 
$\omega_\mathrm{ph}^\mathrm{min}\simeq(\pi/2)(9\pi/4)^{1/3}N^{-1/3}
\bar{r}_\mathrm{s}^{-2}$ which has the observed size scaling of 
$\omega_\mathrm{sd}$. This estimation of the first peak of the non-interacting 
absorption spectrum (dashed lines in Fig.\ \ref{fig:phtdlsda}) agrees within 
20\,\% with the frequency that is obtained from Eq.\ \ref{eqn:freq_mu2} by only 
keeping the one-body (kinetic energy) component. Once we consider exchange and 
correlation corrections, the previous excitation splits according to its total 
spin. The spin selection rules tell us that the lower frequency appears in the 
absorption spectrum (solid lines in Fig.\ \ref{fig:phtdlsda}). The 
corresponding shift can in principle be extracted within the local density 
functional approximation provided the single-particle wave-functions are known.
 
In order to investigate in more detail the evolution of the single-particle 
excitations into spin modes we vary $\alpha_\mathrm{c}$ in Eq.\ \ref{eqn:dyson}
from the noninteracting particle-hole case $\alpha_\mathrm{c}=0$ to full 
spin-dipole excitation $\alpha_\mathrm{c}=1$. Fig.\ \ref{fig:eph_m} shows the 
evolution of the spin-dipole absorption spectrum with $\alpha_\mathrm{c}$ for 
Na$_\mathrm{34}$. One can see that the structure of the spectrum is not 
modified by the interaction strength. The lowest-frequency peak is always the 
dominating one, and its strength is hardly changed. The whole spectrum is 
red-shifted by the interaction, which is globally attractive in this case.

\begin{figure}
  \includegraphics[width=\figwidth]{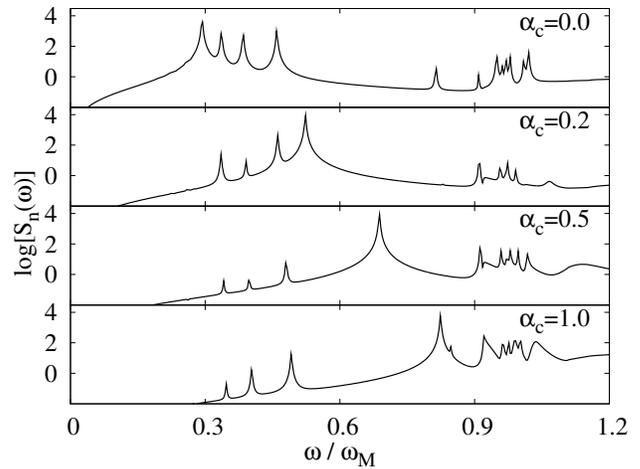}
  \caption{Charge dipole absorption spectrum for $\rm Na_{34}$ and different 
  values of $\alpha_c$. For $\alpha_c=1$, the surface plasmon frequency 
  $\omega_\mathrm{sp} \simeq 0.8\,\omega_\mathrm{M}$ is recovered.}
  \label{fig:eph_n}
\end{figure}
The behavior of the spin dipole can be contrasted with that of the charge 
dipole (Fig.~\ref{fig:eph_n}). When $\alpha_\mathrm{c}=0$, the four
particle-hole excitation modes previously obtained can be observed at 
frequencies lower than $0.5\,\omega_\mathrm{M}\approx 1.7\, \mathrm{eV}$. 
By gradually increasing $\alpha_\mathrm{c}$, three of the modes are slightly 
blue-shifted and become considerably weaker, while the fourth one experiences 
a much larger blue shift and dominates the other peaks for 
$\alpha_\mathrm{c}>0.2$, eventually by several orders of magnitude. 
For $\alpha_\mathrm{c}=1$, this peak coincides with the surface plasmon, 
which is a collective excitation with 
$\omega_{\mathrm sp} \simeq 2.8\, \mathrm{eV}\approx 0.8\,\omega_\mathrm{M}$.

These findings are at odds with the claims of Serra \textit{et al.} 
\cite{serra93}, who interpreted the surface paramagnon as a collective 
excitation. In contrast, our results indicate that the various spin-dipole 
modes appearing in the absorption spectrum should be viewed as individual 
particle-hole excitations, slightly modified by the electron-electron 
interaction.

\section{Open-shell systems} 
\label{sec:open}

For closed-shell systems, an electric dipole field only couples to the charge 
dipole in the linear regime and the excitation of the surface plasmon 
dominates the absorption of laser light. One possibility to observe 
cross-talking between the charge and dipole modes is to operate in the 
nonlinear regime with strong excitations \cite{mornas96}. This may raise some 
practical difficulties, such as electrons escaping the nanoparticle, thus 
leaving behind a net positive charge. Another possibility is to work with 
open-shell systems that possess an intrinsic magnetization in the ground 
state, so that the charge and spin modes are coupled even in the linear regime
\cite{torres00,serra99}. 

\begin{figure}
  \includegraphics[width=\figwidth]{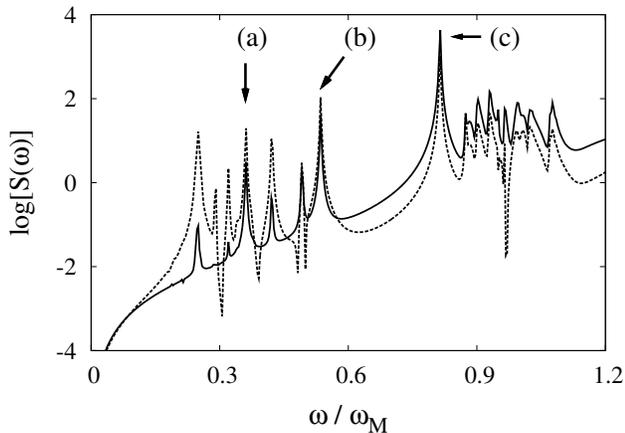}
  \caption{Dipole absorption spectrum $S_\mathrm{n}$ (solid line) and 
    dipole-induced spin-dipole absorption spectrum $S_\mathrm{mn}$ 
    (dashed line) for Na$_\mathrm{27}$. The frequencies of peaks (a), 
    (b), and (c) are $1.22$, $1.82$, and $2.77$ eV, respectively.}
  \label{fig:Na27}
\end{figure}
\begin{figure}
  \includegraphics[width=\figwidth]{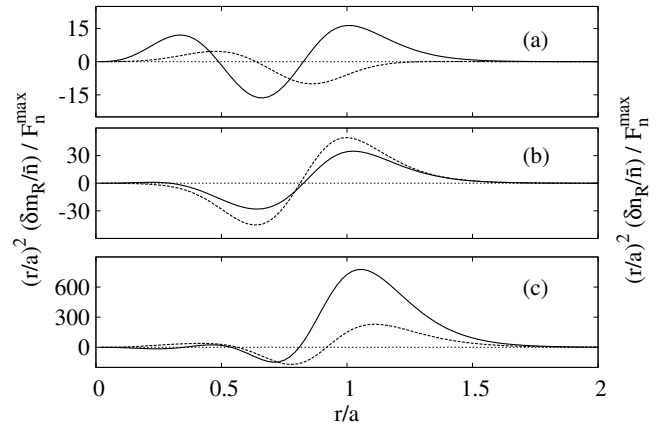}
  \caption{Radial part of the dynamical charge (solid lines) and magnetization 
  densities (dashed lines) corresponding to the peaks highlighted in Fig.\ 
  \ref{fig:Na27}. The normalizations are as in Fig.\ \ref{fig:overviewNa34}.}
  \label{fig:Na27_mag}
\end{figure}

In Fig.\ \ref{fig:Na27}, we show the dipole absorption spectrum $S_\mathrm{n}$ 
and dipole-induced spin-dipole absorption spectrum $S_\mathrm{mn}$ for a 
Na$_\mathrm{27}$ nanoparticle \cite{na27}. In both cases, the system is excited 
by an oscillating electric field which induces both a charge-dipole mode 
(solid curve) and a spin-dipole mode (dotted curve). The charge and 
magnetization profiles corresponding to some of the observed peaks are 
plotted in Fig.\ \ref{fig:Na27_mag}.

The strongest coupling between the charge and spin channels occurs at the
frequency of the surface plasmon. For high-energy peaks, the charge-dipole 
response dominates its spin-dipole counterpart, while the spin-dipole response 
is more important in the low-energy spectrum. The peak labeled (b) in Fig.\ 
\ref{fig:Na27} appears as a special case -- with an energy intermediate between
that of the surface paramagnon and surface plasmon modes -- for which the 
spin-dipole response is comparable to the charge-dipole response. The same 
qualitative features have been observed in the spectra of other open-shell 
systems.

In the sequel we focus on the maximum of the spin-dipole response that occurs 
at the surface plasmon frequency for systems with non-zero ground state 
magnetization (peak (c) in Fig.\ \ref{fig:Na27}). For weak excitations, the 
dynamical magnetization can be written as
\begin{equation}
  \Delta m(\mathbf{r}) = \Delta \xi(\mathbf{r}) n_0(\mathbf{r}) + 
  \Delta n(\mathbf{r}) \xi_0(\mathbf{r})
  \label{eqn:openm1}
\end{equation}
in terms of the ground-state electron density $n_0(\mathbf{r})$, the 
ground-state polarization $\xi_0(\mathbf{r})$, and their dynamical counterparts
$\Delta n(\mathbf{r})$ and $\Delta \xi(\mathbf{r})$. As the laser light 
couples essentially to the charge degrees of freedom, the excitation has the 
form of a shift $Z$ of the entire electron population. The dynamical 
excitation density is thus concentrated at the surface and, in the hard-wall 
homogeneous density approximation, given by 
$\Delta n(\mathbf{r})= \bar{n}Z\delta(a-r)\cos\theta$, just like in the case 
of the surface plasmon. The resulting charge excitation corresponds to a peak 
at the surface plasmon frequency $\omega_\mathrm{sp}$ in the absorption 
spectrum of Fig.\ \ref{fig:Na27}.

\begin{figure}
  \includegraphics[width=\figwidth]{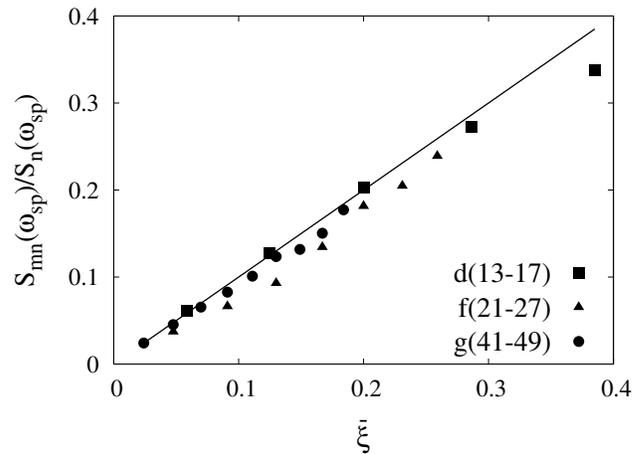}
  \caption{Ratio of the spin-dipole to charge-dipole absorption cross-section 
  for a spin-independent excitation at the surface plasmon frequency, as a 
  function of the mean ground-state spin polarization $\bar{\xi}$. Squares 
  refer to particles Na$_\mathrm{13}$ -- Na$_\mathrm{17}$, triangles to 
  particles Na$_\mathrm{21}$ -- Na$_\mathrm{27}$, and dots to particles 
  Na$_\mathrm{41}$ -- Na$_\mathrm{49}$, where the open electronic shells are 
  d, f, and g, respectively.}
  \label{fig:openxi}
\end{figure}
In addition, even though the excitation does not act directly on the spin 
polarization (\textit{i.e.} $\Delta \xi(\mathbf{r})= 0$), the dynamical 
magnetization corresponding to the charge displacement
$\Delta m(\mathbf{r}) = \Delta n(\mathbf{r}) \xi_0(\mathbf{r})$ does not 
vanish when $\xi_0 \neq 0$. Assuming that the polarization $\xi_0(\mathbf{r})$ 
is uniform inside the particle, and equal to the mean spin polarization 
$\bar{\xi}=(N^\uparrow-N^\downarrow)/N$ (for the example of Na$_{27}$ \cite{na27} 
one has $\bar{\xi}=7/27\approx 0.26$), we obtain a response to the 
spin-independent excitation $V_\mathrm{ex,n}$ in the magnetization channel 
$\Delta m(\mathbf{r})$ which is proportional to the induced 
$\Delta n(\mathbf{r})$ in the charge channel, with proportionality constant 
$\bar{\xi}$. Therefore, we expect
\begin{equation}
 S_\mathrm{mn}(\omega_\mathrm{sp})=\bar{\xi}S_\mathrm{n}(\omega_\mathrm{sp})\, . 
 \label{eqn:openxi2}
\end{equation}
In Fig. \ref{fig:openxi} we present results of TDLSDA calculations with 
spin-independent excitation for the mode with $\omega=\omega_\mathrm{sp}$, for 
a variety of open-shell nanoparticles. The ratio between the spin-dipole and 
charge-dipole absorption cross-sections for not too large polarization is 
indeed to a very good approximation given by the mean ground-state 
polarization $\bar{\xi}$. This allows us to predict particularly strong 
cross-talk between the spin-dipole and the surface plasmon modes for 
open-shell nanoparticles having large ground-state polarizations.  

\section{Conclusion}
\label{sec:conclusions}

In this work, we have studied the spin-dependent linear response in 
alkali-metal (particularly sodium) nanoparticles. Our primary aim was to 
achieve some insight into the nature of these modes, which were first 
investigated by Serra and collaborators \cite{serra93,serra97}. Towards this 
goal we derived simple analytical and semi-analytical models that were 
confronted with linear response TDLSDA calculations. 

The spin-dipole absorption spectrum displays a number of peaks at frequencies 
lower than the surface plasmon frequency. The lowest of them is characterized 
by a magnetization profile without nodes. An excess of spin-up electrons is 
built in half of the nanoparticle at the expense of the spin-down electrons, 
which are majority in the other half of the nanoparticle. The restoring force 
of such a non-equilibrium configuration results in the out-of-phase 
oscillation of the two spin subsystems. The local spin-density approximation 
can be used to estimate the restoring force, and within a classical picture, 
we could estimate the lowest frequency. This approach provides the correct 
scaling of the frequency with the particle size (as $N^{-1/3}$), albeit 
blue-shifted with respect to the TDLSDA results. Such a deviation is partially 
corrected by including the spill-out effect in a phenomenological way using 
the numerically obtained surface plasmon frequencies instead of the Mie value. 
A more sophisticated model, taking into account the inhomogeneities in the 
ground-state density and gradient corrections, yielded an even better 
agreement.

By comparing the spin-dipole absorption spectrum with that obtained by 
progressively removing the electron-electron interaction, we observed a 
one-to-one correspondence of the particle-hole excitations and the spin-dipole 
modes. We thus showed that the spin-dipole modes are slight perturbations of 
the particle-hole excitations, and therefore do not qualify as genuine 
collective excitations, contrary to the claim of Ref.~\onlinecite{serra93}.

Finally, we studied the possibility of exciting the spin-dipole modes by 
ordinary optical means (laser pulses). For open-shell systems, it is 
well-known that the spin and charge modes are coupled in the linear regime 
\cite{torres00,serra99}. We showed that, when exciting the system with a 
density shift, a spin-dipole mode appears at the surface plasmon frequency, 
together with the standard charge-dipole mode. The ratio between the strengths 
of the absorption peak for the spin dipole and charge dipole modes was shown 
to be given by the spin polarization of the ground-state.

While our numerical calculations were done in the case of Na nanoparticles,
our general conclusions are also valid for noble-metal nanoparticles and 
semiconductor quantum dots. The latter systems are more adapted than alkaline 
nanoparticles for experimental spectroscopic studies. Moreover, the concepts 
developed for the study of spin modes in normal-metal nanoparticles could be 
useful in analyzing ferromagnetic nanoparticles, in view of the strong 
internal field existing in these materials. 

\begin{acknowledgments}
We thank M.\ Barranco for helpful correspondence and we acknowledge financial 
support from the French National Research Agency ANR (project ANR-06-BLAN-0059).
\end{acknowledgments}

\appendix

\section{LSDA parametrization}
\label{sec:app-perdew}

Our numerical and analytical approaches to obtain the spin-dipole resonances
are based on the local spin density approximation \cite{lipparini-book}.
For completeness we present in this appendix the particular parametrization
that we chose in our approaches. The energy functional of the
electron system can be written as
\begin{equation}\label{eqn:eng1}
  E [n^\uparrow, n^\downarrow] = E_\mathrm{K}[n^\uparrow, n^\downarrow]
             + E_\mathrm{H}[n]+ E_\mathrm{XC}[n^\uparrow,n^\downarrow]\, ,
\end{equation}
where $E_\mathrm{K}$ represents the kinetic energy, $E_\mathrm{H}$ the
Hartree contribution, and $E_\mathrm{XC}$ the exchange-correlation term.
The kinetic energy is given by
\begin{equation}\label{eqn:ekingrad}
  E_\mathrm{K}[n^\uparrow, n^\downarrow] =
  E_\mathrm{K,TF}[n^\uparrow, n^\downarrow]+
  E_\mathrm{K,G}[n^\uparrow, n^\downarrow] ,
\end{equation}
with the Thomas-Fermi component
\begin{equation}\label{eqn:ekintf}
 E_\mathrm{K,TF}[n^\uparrow, n^\downarrow] =
 \frac{3}{10} \left(6\pi^2\right)^{2/3}
 \int \mathrm{d}\mathbf{r}
  \left(n^{\uparrow 5/3}(\mathbf{r}) + n^{\downarrow 5/3}(\mathbf{r})\right)
\end{equation}
and the gradient correction
\begin{equation}\label{eqn:ekingrad_2}
  E_\mathrm{K,G}[n^\uparrow, n^\downarrow] =
  \frac{1}{72} \int \mathrm{d}\mathbf{r}
         \left( \frac{|\nabla n^\uparrow |^2}{n^\uparrow}
              + \frac{|\nabla n^\downarrow|^2}{n^\downarrow}\right)\, ,
\end{equation}
which takes into account the spatial density variations in the electron gas 
\cite{Oliver79,brack93}. This term is particularly relevant when we consider 
spill-out effects and thus a finite region where the density exhibits large 
spatial variations (as in Sec.~\ref{subsec:nued}).

The Hartree contribution depends only on the total density
$n=n^\uparrow+n^\downarrow$, and is given by the electrostatic potential energy
\begin{equation}  \label{eq:Hartree}
 E_\mathrm{H}[n^\uparrow, n^\downarrow] =
  \frac{1}{2} \iint \mathrm{d}\mathbf{r}\mathrm{d}\mathbf{r}'\,
    \frac{n(\mathbf{r})n(\mathbf{r}')}{\left|\mathbf{r}-\mathbf{r}'\right|}\, .
\end{equation}
This term is irrelevant for spin-dipole excitations in spherically symmetric 
nanoparticles since they do not involve charge displacements.

The exchange-correlation term can be expressed in terms of the exchange and 
correlation components as $E_\mathrm{XC}=E_\mathrm{X}+E_\mathrm{C}$ with
\begin{equation}\label{eq:exc}
E_\mathrm{X/C}[n^\uparrow, n^\downarrow] =
  \int \mathrm{d}\mathbf{r} \, n(\mathbf{r})
     \epsilon_\mathrm{x/c}(n^\uparrow, n^\downarrow)\, .
\end{equation}
The function $\epsilon_\mathrm{xc}=\epsilon_\mathrm{x}+\epsilon_\mathrm{c}$
determines the exchange-correlation potential through Eq.\ \ref{eqn:exch-corr}. 
For the exchange part one has
\begin{equation}
  \label{eq:ex}
    \epsilon_\mathrm{x}(n^\uparrow, n^\downarrow) =
    - \frac{3}{2} \left(\frac{3}{4 \pi}\right)^{1/3}
       \left(n^{\uparrow 4/3}+ n^{\downarrow 4/3}\right)/n\, .
\end{equation}
For $\epsilon_\mathrm{c}$ we use \cite{perdew81}
\begin{equation}
\epsilon_\mathrm{c} \left(n^\uparrow, n^\downarrow\right) =
  \epsilon_\mathrm{c}^\mathrm{U}(r_\mathrm{s})
   + \left[ \epsilon_\mathrm{c}^\mathrm{P} (r_\mathrm{s}) -
            \epsilon_\mathrm{c}^\mathrm{U} (r_\mathrm{s})\right] f(\xi)
\end{equation}
with
\begin{eqnarray}
 \epsilon_\mathrm{c}^\mathrm{U}(r_\mathrm{s}) & = &
     \frac{-0.1423}{1 + 1.0529 \sqrt{r_\mathrm{s}} + 0.3334\, r_\mathrm{s}} \\
 \epsilon_\mathrm{c}^\mathrm{P}(r_\mathrm{s}) & = &
     \frac{-0.0843}{1 + 1.3981 \sqrt{r_\mathrm{s}} + 0.2611\, r_\mathrm{s}} \\
 \label{eq:def_f}
 f(\xi) & = & \frac{(1+\xi)^{4/3} + (1-\xi)^{4/3}-2}{2^{4/3}-2}\, .
\end{eqnarray}
The normalized inter-particle distance $r_\mathrm{s}=(4\pi n/3)^{-1/3}$ and the 
spin polarization $\xi=(n^\uparrow-n^\downarrow)/n$ are both local properties 
of the electron system. The above parametrization of $E_\mathrm{XC}$ was used 
in our analytical approaches, as well as in the numerics, since it has been 
proven to provide a good representation for the electron densities that we 
are interested in.

\section{Magnetization profile for non-uniform ground-state densities}
\label{sec:app-calc_nu}

In this appendix we develop the perturbed equilibrium condition 
\eqref{eq:magnet} for a closed-shell nanoparticle in an external magnetic 
field and derive the differential equation \eqref{eqn:in_mag5} for the 
magnetization profile. 

Expressing Eq.\ \ref{eq:magnet} in terms of the polarization we have 
\begin{equation}
\frac{1}{n_0(r)} \left. \frac{\delta E[\xi]}
{\delta \xi(\mathbf{r})}\right|_{\xi=\Delta \xi(\mathbf{r})} = 
\frac{z}{\lambda_B} \, . 
\label{eq:aa1}
\end{equation}
Similarly as in \eqref{eq:energy_changes}, we can write the functional
derivatives of $E_\mathrm{K,TF}$, $E_\mathrm{X}$, and $E_\mathrm{C}$, 
respectively, as
\begin{widetext}
\begin{subequations}
\label{eq:aa2}
\begin{eqnarray}
  \frac{1}{n_0(r)}\frac{\delta E_{\mathrm{K,TF}}}{\delta \xi(\mathbf{r})} &=&
    \frac{(3\pi^2)^{2/3}}{4} \ n_0^{2/3}(r) \ \left(
   \left[1+\xi(\mathbf{r})\right]^{2/3}-\left[1-\xi(\mathbf{r})\right]^{2/3}
   \right)\, , \\
   \frac{1}{n_0(r)}\frac{\delta E_{\mathrm{X}}}{\delta \xi(\mathbf{r})} &=&
   -\frac{1}{2}\left(\frac{3}{\pi}\right)^{1/3} n_0^{1/3}(r) \left( 
   \left[1+\xi(\mathbf{r})\right]^{1/3}-\left[1-\xi(\mathbf{r})\right]^{1/3}
   \right) \, , \\
   \frac{1}{n_0(r)}\frac{\delta E_{\mathrm{C}}}{\delta \xi(\mathbf{r})} &=&
    \left[\epsilon_\mathrm{c}^\mathrm{P}\left(r_\mathrm{s}(r)\right)-
   \epsilon_\mathrm{c}^\mathrm{U}\left(r_\mathrm{s}(r)\right)\right] 
    f'(\xi(\mathbf{r})) \, , 
\end{eqnarray}
\end{subequations}
where $f'$ stands for the derivative of the function $f$ defined in 
\eqref{eq:def_f}. Because of large density variations in the spill-out region 
we consider the gradient correction \eqref{eqn:ekingrad_2} for which we get
\begin{equation}
\frac{1}{n_0(r)}\frac{\delta E_{\mathrm{K,G}}}{\delta \xi(\mathbf{r})} =
   \frac{1}{72 n_0(r)} \frac{\delta}{\delta \xi(\mathbf{r})}
    \left\{\int \mathrm{d}\mathbf{r}
         \left(\frac{|\nabla n_0|^2}{n_0(r)} + n_0(r)
   \frac{|\nabla \xi(\mathbf{r})|^2}{1-\xi^2(\mathbf{r})}\right)\right\}\, .
   \end{equation}
The first term of the integrand is independent of $\xi(\mathbf{r})$, and 
the second can be treated using partial integration yielding
\begin{equation}
\frac{1}{n_0(r)}\frac{\delta E_{\mathrm{K,G}}}{\delta \xi(\mathbf{r})} =
   \frac{1}{36}\left(
   \frac{\xi(\mathbf{r})|\nabla \xi(\mathbf{r})|^2}{[1-\xi^2(\mathbf{r})]^2}-
   \frac{1}{n_0(r)}\nabla\left(\frac{n_0(r)\nabla \xi(\mathbf{r})}
   {1-\xi^2(\mathbf{r})}\right)\right)  \, .
\end{equation}
\end{widetext}
Since $\nabla n_0(r)=n_0'(r)\hat{\mathbf{r}}$, we have 
$\nabla(n_0(r)\nabla \xi(\mathbf{r}))=
n_0'(r)(\partial \xi(\mathbf{r})/\partial r)+ n_0(r)\nabla^2\xi(\mathbf{r})$.
Assuming that the polarization is small and a smooth function of $\mathbf{r}$ 
verifying $\nabla^2\xi \ll \partial \xi(\mathbf{r})/\partial r$ we remain 
in linear order in $\xi$ and write
\begin{equation}
\frac{1}{n_0(r)}\frac{\delta E_{\mathrm{K,G}}}{\delta \xi(\mathbf{r})} \approx
   - \frac{1}{36} \ \frac{n_0'(r)}{n_0(r)} \ 
   \frac{\partial \xi(\mathbf{r})}{\partial r} \, .
\end{equation}
Gathering the various contributions to the energy, Eq.~\ref{eq:aa1} becomes  
\begin{equation} \label{eq:aa3}
   -\frac{1}{36} \ \frac{n_0'(r)}{n_0(r)} \ 
   \frac{\partial \Delta\xi(\mathbf{r})}{\partial r} 
   + A_\mathrm{KS} \xi(\mathbf{r})= \frac{r}{\lambda_B} \cos{\theta} \, ,
\end{equation}
with $A_\mathrm{KS}$ defined in \eqref{eq:def_dif_eq_a}. Since Eq.~\ref{eq:aa3}
admits solutions of the dipolar form, we write $\Delta\xi(\mathbf{r})=
\delta\xi_\mathrm{R}(r)\cos{\theta}$ and $\tilde{m}(r)=
\lambda_B r^2 n(r)\delta\xi_\mathrm{R}(r)$, obtaining Eq.~\ref{eqn:in_mag5} 
for the magnetization profile. The function $\tilde{m}(r)$ is more appropriate 
than $\delta\xi_\mathrm{R}(r)$ for numerical calculations dealing with small 
electron densities.


\begin{thebibliography}{99}
\bibitem{na34}
  The electron configuration of $\rm Na_{34}$ can be expressed as
  $1s^2$ $1p^6$ $1d^{10}$ $2s^2$ $1f^{14}$.
\bibitem{note-sp}
  Following similar lines in the case of a charge-dipole excitation
  we obtain for the surface plasmon frequency the estimation
  $\hat\omega_\mathrm{M}^2 = (3N/4\pi)(\int \mathrm{d}r \, r \tilde{n}(r))^{-1}$,
  where $\tilde{n}(r)=\lambda_E r^2 \delta n_\mathrm{R}(r)$ with
  $\delta n_\mathrm{R}(r)$ being the radial part of the density
  perturbation and $\lambda_E$ giving the strength of the perturbing field.
  Analogously to Eq.~\ref{eqn:in_mag5}, $\tilde{n}$ verifies an
  integro-differential equation involving a Hartree term, as well as the
  other contributions to the energy \eqref{eqn:eng1}.
\bibitem{na27}
  The electron configuration of $\rm Na_{27}$ can be expressed as
  $1s^2$ $1p^6$ $1d^{10}$ $2s^2$ $1f^{7}$. According to Hund's rule, 
  the spins of  the 7 $f$-electrons are aligned.
\end{thebibliography}
\end{document}